\documentclass[10pt, sigconf]{acmart}

\usepackage{booktabs} 
\usepackage{float}

\setcopyright{none}

\begin{document}
\title[A Deep-RL Approach for SDN Routing Optimization]{A Deep-Reinforcement Learning Approach for Software-Defined Networking Routing Optimization}

\author{Giorgio Stampa*, Marta Arias*, David S\'anchez-Charles**, Victor Munt\'es-Mulero**, Albert Cabellos*}
\affiliation{%
  \institution{* Universitat Polit\`ecnica de Catalunya, ** CA Technologies}
  \institution{\{giorgio,acabello\}@ac.upc.edu, marias@cs.upc.edu, \{david.sanchez,victor.muntes\}@ca.com}
}





\renewcommand{\shortauthors}{Stampa et al.}

\begin{abstract}
In this paper we design and evaluate a Deep-Reinforcement Learning agent that optimizes routing. 
Our agent adapts automatically to current traffic conditions and proposes tailored configurations that attempt to minimize the network delay.
%
Experiments show very promising performance.
Moreover, this approach provides important operational advantages with respect to traditional optimization algorithms.
\end{abstract}

%
%




\maketitle

\section{Introduction}

Recent trends in networking point to the use of Machine Learning (ML) techniques for the control and operation of the network. Although the concept in itself is not new~\cite{Clark03}, these trends are gaining momentum thanks to two enabling technologies: Software-Defined Networking (SDN) ~\cite{Kreutz15} and Network Analytics (NA) ~\cite{Kim15,Clemm15}. This new networking paradigm is referred to as Knowledge-Defined Networking (KDN) ~\cite{Mestres17}.

%
%

In this paper we focus on the use of a Deep-Reinforcement Learning (DRL) agent for routing optimization. By taking advantage of the recent breakthroughs of deep neural networks applied to reinforcement learning ~\cite{Li17,Arulkumaran17} we design and train a DRL agent capable of optimizing routing according to a predefined target metric: network delay. As we show, our DRL agent provides promising performance against an initial benchmark while providing important operational advantages.




\section{State of the art}

Routing optimization and particularly traffic engineering (TE) are well-established topics in networking whose goal is to control the behavior of transmitted data in order to maximize the performance of the network, according to targets defined by the operator. A wide range of solutions exist, generally based on analytical optimization or on local-search heuristic methods ~\cite{Wang08}. 

The rise of the SDN paradigm has resulted in additional degrees of freedom in routing (flow- vs. destination-based forwarding). This has increased the size of the solution space as well as the optimization capabilities and as a result a plethora of new solutions are being proposed ~\cite{Akyildiz14}.

In the area of ML, RL techniques have already been used in the context of routing optimization, which was pioneered in ~\cite{Boyan94}. Recent attempts also use RL techniques, in this case to achieve QoS routing ~\cite{Lin16}. Such proposals use table-based RL agents, which during training fill a table \([state \times action] \rightarrow reward\) that is then used while in operation.




In this context, we propose to improve with respect to the state of the art by: (i) being able to generalize over and provide solutions for unseen network states, which cannot be achieved by traditional table-based RL agents; (ii) overcoming the iterative improvement steps of optimization and heuristics by having a DRL agent provide a near-optimal solution in one single step (after training, either on- or off-line). To the best of our knowledge this is the first attempt to use DRL techniques for routing optimization.

\section{Deep RL Agent}

The proposed RL agent is an off-policy, actor-critic, deterministic policy gradient algorithm ~\cite{Silver14} that interacts with its environment (i.e. the network) through the exchange of the three signals ~\textit{state}, ~\textit{action} and ~\textit{reward} ~\cite{Sutton98}.


Specifically, the state is represented by the Traffic Matrix (TM, being the bandwidth request between each source-destination pair), the action by a tuple of link-weights (that univocally determine the paths for all source-destination node pairs), and the reward is based on the mean network delay.

%
%

The objective of the agent is to determine the optimal behavior policy \(\pi\) mapping from the space of states \(S\) to the space of actions \(A\) (\(\pi: S \rightarrow A\)) that maximizes the expected reward \(r \in \mathbb{R}\) (minimizes the network delay). It does so by iteratively improving its knowledge of the relationship between the three signals by means of two deep neural networks (one for the ~\textit{actor}, one for the ~\textit{critic}) ~\cite{Lillicrap15}.

\section{Experimental Results}

\textbf{Methodology:} To assess the performance of the proposed DRL agent we use a scale-free network topology of 14 nodes and 21 full-duplex links, with uniform link capacities and average node degree of 3. We use 10 traffic intensity levels (TI), ranging from 12.5\% to 125\% of the total network capacity. For each TI we generate 100 TMs of equal total traffic using a gravity model ~\cite{Roughan05}, therefore having 1,000 distinct traffic configurations varying over both the total volume of traffic and its distribution.

As an initial step towards evaluating our DRL agent, we compare its performance against 100,000 
randomly-generated routing configurations.
These configurations are valid in the sense that: (i) all nodes are reachable, (ii) there are no loops.
For each combination of TI and TM the same 100,000 routing configurations are used.

The DRL agent was trained for 100,000 steps with gravity-generated TMs. Routing configurations (i.e. actions) were determined by the agent itself. We added a stochastic exploration rate to the agent's actions (overriding its deterministic action choices) to avoid it getting stuck in local minima during training ~\cite{Sutton98}. We then presented each of the 1,000 test TMs to the trained DRL agent to obtain its predicted solution in one single step.

The OMNeT++ discrete event simulator (v4.6) ~\cite{Varga01,Varga08} was used throughout the experiment to obtain the network delay under given traffic and routing. All datasets and source code used in this paper are available at ~\cite{kdn}.

\begin{figure}
\includegraphics[width=3in,height=2in]{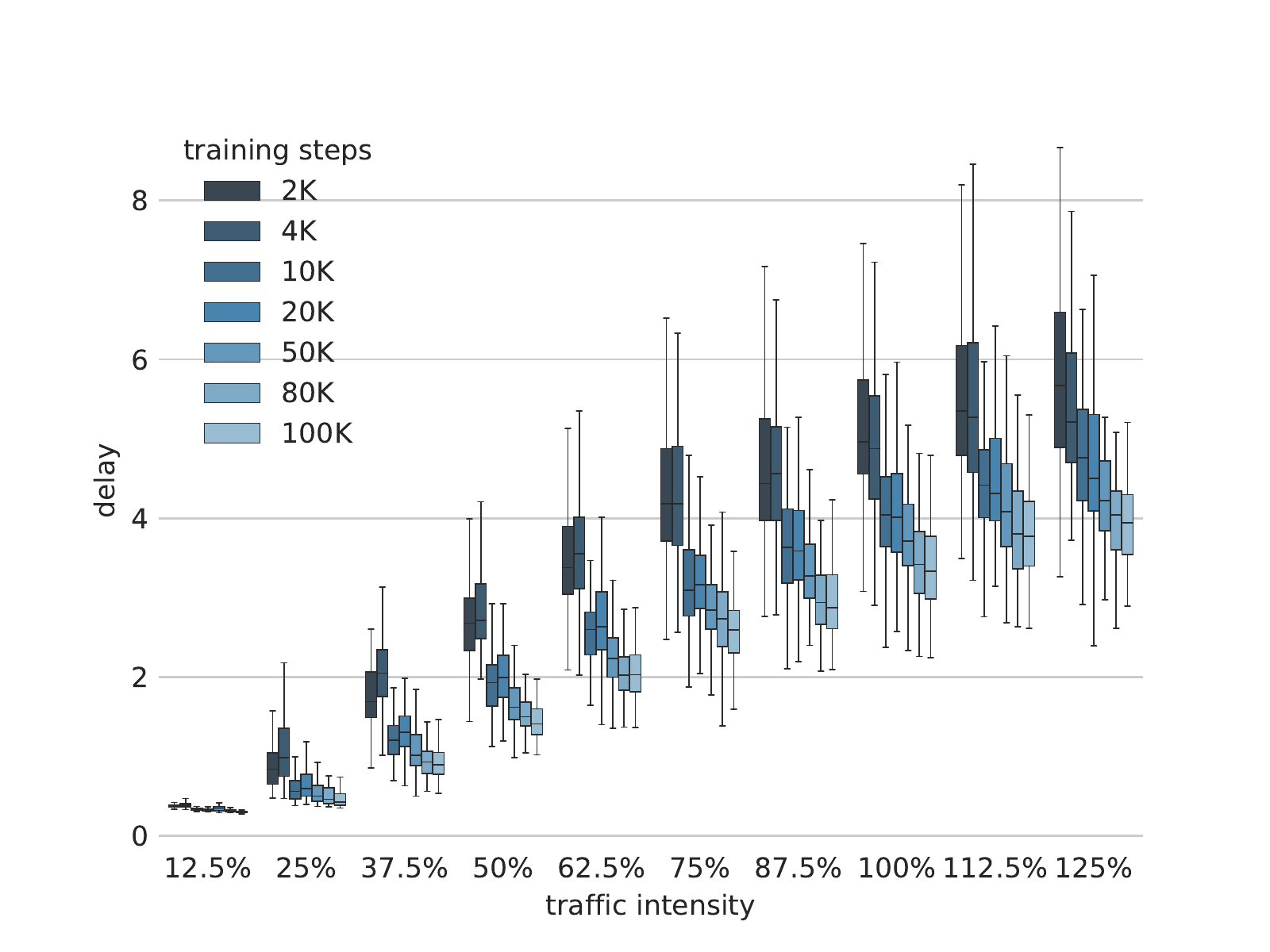}
\caption{DRL learning progress.}
\label{fig:training}
\end{figure}

\begin{figure}
\includegraphics[width=3in,height=2in]{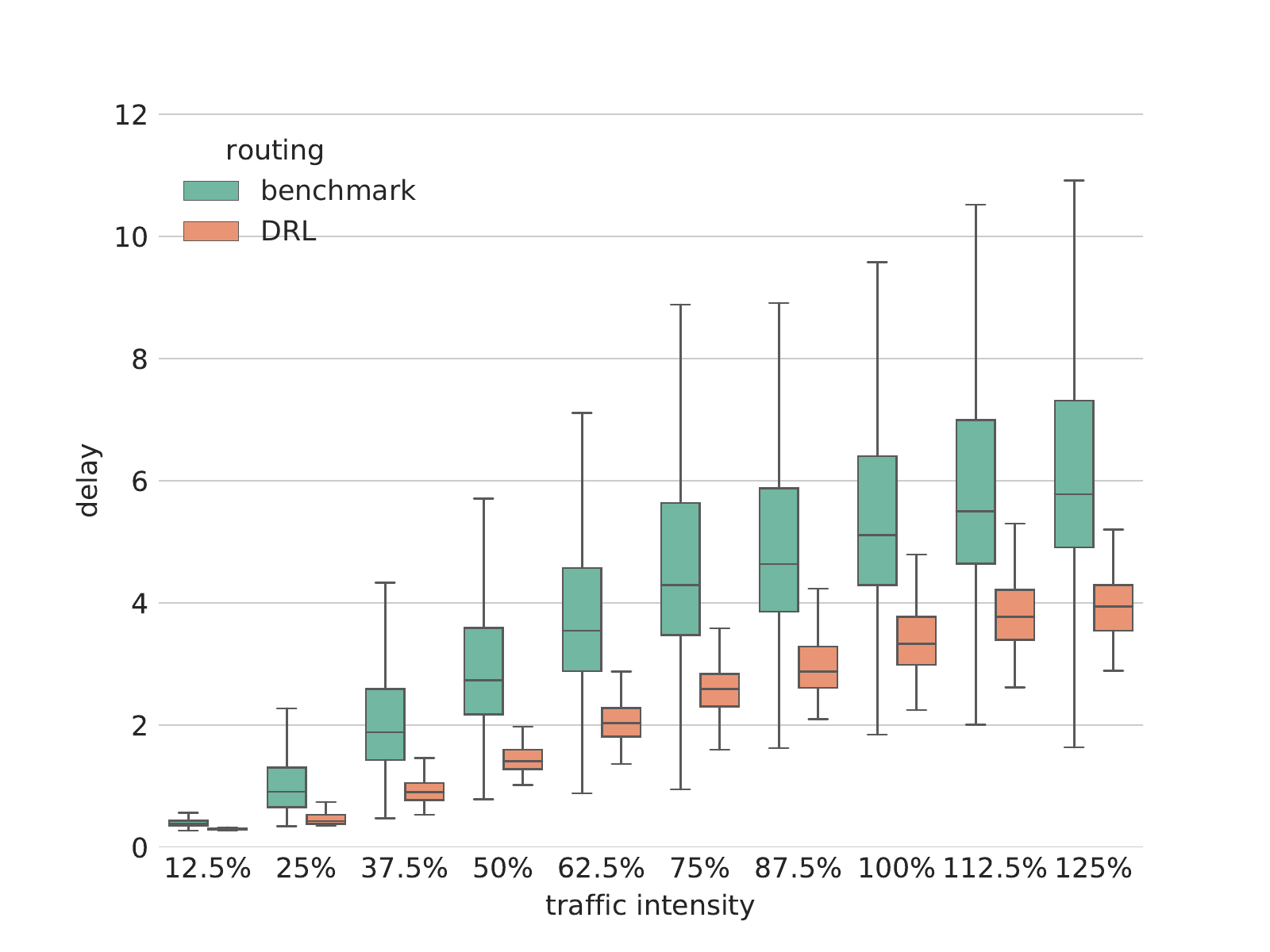}
\caption{DRL vs. benchmark.}
\label{fig:comparison}
\end{figure}


\textbf{Results:} Figures \ref{fig:training} and \ref{fig:comparison} show the performance improvement of the DRL agent over training episodes and how it compares with our benchmark, respectively. Simulations are grouped per TI (x-axis); different strategies for routing determination are shown side-by-side. The bottom and the top of the box plots represent the 1st and 3rd quartile of data and the whisker lines the lowest/highest observation within 1.5 * IQR (interquartile range) of the lower/upper quartile. Outliers are not shown for the sake of clarity.

The first relevant outcome is that the DRL agent's performance increases with training time, as is reasonably expected (Fig.~\ref{fig:training}). The second is that our trained agent consistently computes, over all TIs, routing configurations that are, on average, well within the 1st quartile of our benchmark (Fig.~\ref{fig:comparison}).

%
%

\section{Discussion}

In this paper we have shown that a fully automated DRL agent can provide routing configurations that tend to minimize the network's delay. In the immediate future, we plan on expanding our experiments to include more sophisticated benchmarks and topologies for routing optimization.

%
%

We find several advantages of DRL architectures over traditional heuristic and linear/convex optimization techniques:


\textbf{One-step optimization:} Once trained, the DRL agent can produce a near-optimal routing configuration in one single step. On the contrary, traditional optimization requires a large number of steps to produce a new configuration. This represents an important advantage for real-time control of the network.

\textbf{Model-free:} DRL agents are model-free (they learn autonomously from experience the dynamics between ~\textit{state}, ~\textit{action} and ~\textit{reward} ~\cite{Sutton98}) and can understand non-linear, complex, multi-dimensional systems, without making any simplifications. On the other hand, traditional optimization algorithms require an analytical model of the underlying system deriving from significant assumptions and simplifications.

\textbf{Black-box optimization:} An appealing advantage of DRL agents is that they assume the system being a black-box and are automatic. Traditional mechanisms, particularly heuristics, are tailored for the problem they are trying to optimize. With DRL agents, different reward functions can be used to implement different target policies, without the need of designing a new algorithm.



\bibliographystyle{unsrtnat}

\begin{thebibliography}{18}
\providecommand{\natexlab}[1]{#1}
\providecommand{\url}[1]{\texttt{#1}}
\expandafter\ifx\csname urlstyle\endcsname\relax
  \providecommand{\doi}[1]{doi: #1}\else
  \providecommand{\doi}{doi: \begingroup \urlstyle{rm}\Url}\fi

\bibitem[Clark et~al.(2003)Clark, Partridge, Ramming, and Wroclawski]{Clark03}
David~D. Clark, Craig Partridge, J.~Christopher Ramming, and John~T.
  Wroclawski.
\newblock A knowledge plane for the internet.
\newblock In \emph{Proceedings of the 2003 Conference on Applications,
  Technologies, Architectures, and Protocols for Computer Communications},
  SIGCOMM '03, pages 3--10, New York, NY, USA, 2003. ACM.
\newblock ISBN 1-58113-735-4.

\bibitem[Kreutz et~al.(2015)Kreutz, Ramos, Verissimo, Rothenberg, Azodolmolky,
  and Uhlig]{Kreutz15}
Diego Kreutz, Fernando~MV Ramos, Paulo~Esteves Verissimo, Christian~Esteve
  Rothenberg, Siamak Azodolmolky, and Steve Uhlig.
\newblock Software-defined networking: A comprehensive survey.
\newblock \emph{Proceedings of the IEEE}, 103\penalty0 (1):\penalty0 14--76,
  2015.

\bibitem[Kim et~al.(2015)Kim, Sivaraman, Katta, Bas, Dixit, and Wobker]{Kim15}
Changhoon Kim, Anirudh Sivaraman, Naga Katta, Antonin Bas, Advait Dixit, and
  Lawrence~J Wobker.
\newblock In-band network telemetry via programmable dataplanes.
\newblock In \emph{ACM SIGCOMM}, 2015.

\bibitem[Clemm et~al.(2015)Clemm, Chandramouli, and Krishnamurthy]{Clemm15}
Alexander Clemm, Mouli Chandramouli, and Sailesh Krishnamurthy.
\newblock Dna: An sdn framework for distributed network analytics.
\newblock In \emph{Integrated Network Management (IM), 2015 IFIP/IEEE
  International Symposium on}, pages 9--17. IEEE, 2015.

\bibitem[Mestres et~al.(2017)Mestres, Rodriguez-Natal, Carner, Barlet-Ros,
  Alarc\'{o}n, Sol{\'e}, Munt{\'e}s-Mulero, Meyer, Barkai, Hibbett, Estrada,
  Ma'ruf, Coras, Ermagan, Latapie, Cassar, Evans, Maino, Walrand, and
  Cabellos]{Mestres17}
Albert Mestres, Alberto Rodriguez-Natal, Josep Carner, Pere Barlet-Ros, Eduard
  Alarc\'{o}n, Marc Sol{\'e}, Victor Munt{\'e}s-Mulero, David Meyer, Sharon
  Barkai, Mike~J. Hibbett, Giovani Estrada, Khaldun Ma'ruf, Florin Coras, Vina
  Ermagan, Hugo Latapie, Chris Cassar, John Evans, Fabio Maino, Jean Walrand,
  and Albert Cabellos.
\newblock Knowledge-defined networking.
\newblock \emph{SIGCOMM Comput. Commun. Rev.}, 47\penalty0 (3):\penalty0 2--10,
  September 2017.
\newblock ISSN 0146-4833.

\bibitem[Li(2017)]{Li17}
Yuxi Li.
\newblock Deep reinforcement learning: An overview.
\newblock \emph{arXiv preprint arXiv:1701.07274}, 2017.

\bibitem[Arulkumaran et~al.(2017)Arulkumaran, Deisenroth, Brundage, and
  Bharath]{Arulkumaran17}
Kai Arulkumaran, Marc~Peter Deisenroth, Miles Brundage, and Anil~Anthony
  Bharath.
\newblock A brief survey of deep reinforcement learning.
\newblock \emph{arXiv preprint arXiv:1708.05866}, 2017.

\bibitem[Wang et~al.(2008)Wang, Ho, Pavlou, and Howarth]{Wang08}
Ning Wang, Kin Ho, George Pavlou, and Michael Howarth.
\newblock An overview of routing optimization for internet traffic engineering.
\newblock \emph{IEEE Communications Surveys \& Tutorials}, 10\penalty0 (1),
  2008.

\bibitem[Akyildiz et~al.(2014)Akyildiz, Lee, Wang, Luo, and Chou]{Akyildiz14}
Ian~F Akyildiz, Ahyoung Lee, Pu~Wang, Min Luo, and Wu~Chou.
\newblock {A roadmap for traffic engineering in SDN-OpenFlow networks}.
\newblock \emph{Computer Networks}, 71\penalty0 (C):\penalty0 1--30, October
  2014.

\bibitem[Boyan and Littman(1994)]{Boyan94}
Justin~A Boyan and Michael~L Littman.
\newblock Packet routing in dynamically changing networks: A reinforcement
  learning approach.
\newblock In \emph{Advances in neural information processing systems}, pages
  671--678, 1994.

\bibitem[Lin et~al.(2016)Lin, Akyildiz, Wang, and Luo]{Lin16}
Shih-Chun Lin, Ian~F Akyildiz, Pu~Wang, and Min Luo.
\newblock {QoS-Aware Adaptive Routing in Multi-layer Hierarchical Software
  Defined Networks: A Reinforcement Learning Approach}.
\newblock In \emph{2016 IEEE International Conference on Services Computing
  (SCC)}, pages 25--33. IEEE, June 2016.

\bibitem[Silver et~al.(2014)Silver, Lever, Heess, Degris, Wierstra, and
  Riedmiller]{Silver14}
David Silver, Guy Lever, Nicolas Heess, Thomas Degris, Daan Wierstra, and
  Martin Riedmiller.
\newblock Deterministic policy gradient algorithms.
\newblock In \emph{Proceedings of the 31st International Conference on Machine
  Learning (ICML-14)}, pages 387--395, 2014.

\bibitem[Sutton and Barto(1998)]{Sutton98}
Richard~S Sutton and Andrew~G Barto.
\newblock \emph{Reinforcement learning: An introduction}.
\newblock MIT press Cambridge, 1998.

\bibitem[Lillicrap et~al.(2015)Lillicrap, Hunt, Pritzel, Heess, Erez, Tassa,
  Silver, and Wierstra]{Lillicrap15}
Timothy~P Lillicrap, Jonathan~J Hunt, Alexander Pritzel, Nicolas Heess, Tom
  Erez, Yuval Tassa, David Silver, and Daan Wierstra.
\newblock Continuous control with deep reinforcement learning.
\newblock \emph{arXiv preprint arXiv:1509.02971}, 2015.

\bibitem[Roughan(2005)]{Roughan05}
Matthew Roughan.
\newblock Simplifying the synthesis of internet traffic matrices.
\newblock \emph{ACM SIGCOMM Computer Communication Review}, 35\penalty0
  (5):\penalty0 93--96, 2005.

\bibitem[Varga(2001)]{Varga01}
Andr{\'a}s Varga.
\newblock Discrete event simulation system.
\newblock In \emph{Proc. of the European Simulation Multiconference
  (ESM'2001)}, 2001.

\bibitem[Varga and Hornig(2008)]{Varga08}
Andr{\'a}s Varga and Rudolf Hornig.
\newblock An overview of the omnet++ simulation environment.
\newblock In \emph{Proceedings of the 1st international conference on
  Simulation tools and techniques for communications, networks and systems \&
  workshops}, page~60. ICST (Institute for Computer Sciences,
  Social-Informatics and Telecommunications Engineering), 2008.

\bibitem[de~Catalunya()]{kdn}
Universitat~Polit\`ecnica de~Catalunya.
\newblock Knowledge-defined networking training datasets.
\newblock URL \url{http://knowledgedefinednetworking.org}.

\end{thebibliography}

\end{document}